\documentclass[twocolumn,pra]{revtex4}
\usepackage[latin9]{inputenc}
\usepackage{float}
\usepackage{amsmath}
\usepackage{amssymb}
\usepackage{wasysym}
\usepackage{graphicx}

\makeatletter
\@ifundefined{textcolor}{}
{%
 \definecolor{BLACK}{gray}{0}
 \definecolor{WHITE}{gray}{1}
 \definecolor{RED}{rgb}{1,0,0}
 \definecolor{GREEN}{rgb}{0,1,0}
 \definecolor{BLUE}{rgb}{0,0,1}
 \definecolor{CYAN}{cmyk}{1,0,0,0}
 \definecolor{MAGENTA}{cmyk}{0,1,0,0}
 \definecolor{YELLOW}{cmyk}{0,0,1,0}
}

\usepackage{amsfonts}

\makeatother

\begin{document}

\title{Brillouin scattering induced transparency and non-reciprocal light
storage}

\author{Chun-Hua Dong$^{1,2}$}

\email{chunhua@ustc.edu.cn}

\author{Zhen Shen$^{1,2}$}

\author{Chang-Ling Zou$^{1,2}$}

\email{clzou321@ustc.edu.cn}

\author{Yan-Lei Zhang$^{1,2}$}

\author{Wei Fu$^{1,2}$}

\author{Guang-Can Guo$^{1,2}$}

\affiliation{$^{1}$Key Laboratory of Quantum Information, University of Science
and Technology of China, Hefei 230026, P. R. China.}

\affiliation{$^{2}$Synergetic Innovation Center of Quantum Information and Quantum
Physics, University of Science and Technology of China, Hefei, Anhui
230026, P. R. China.}

\maketitle
\textbf{Stimulated Brillouin scattering (SBS) is a very fundamental
interaction between light and travelling acoustic waves }\cite{AOP09,AOP13}\textbf{,
which is mainly attributed to the electrostriction and photoelastic
effects with the interaction strength being orders of magnitude larger
than other nonlinearities. Although various photonic applications
for all-optical light controlling based on SBS have been achieved
in optical fiber and waveguides }\cite{Zhu07,Okawachi,Laser,Isolator}\textbf{,
the coherent light-acoustic interaction remains a challenge. Here,
we experimentally demonstrated the Brillouin scattering induced transparency
(BSIT) in a high quality optical microresonantor. }\textbf{\emph{}}\textbf{Benefited
from the triple-resonance in the whispering gallery cavity, the photon-phonon
interaction is enhanced, and enables the light storage to the phonon,
which has lifetime up to $10\ \mathrm{\mu s}$. In addition, due to
the phase matching condition, the stored circulating acoustic phonon
can only interact with certain direction light, which leads to non-reciprocal
light storage and retrieval. Our work paves the way towards the low
power consumption integrated all-optical switching, isolator and circulator,
as well as quantum memory.}

Stimulated Brillouin Scattering (SBS) in fiber and waveguide has stimulated
various photonic applications in past decades \cite{AOP09,AOP13},
such as light storage \cite{Zhu07}, slow light \cite{Okawachi},
laser \cite{Laser} and optical isolator \cite{Isolator}. Recently,
great progresses have been achieved by incorporating the SBS into
photonic integrated chips. On the one hand, the experimentally demonstrated
on-chip SBS \cite{onchip}, where the tightly confinement of fields
in the compact integrated devices, has greatly enhanced the SBS interaction
\cite{Shin13}. And new physics at nanoscale that giant enhancement
of SBS due to radiation pressures or boundary-induced nonlinearities
have also been revealed \cite{Shin13,Rakich}. On the other hand,
the SBS have been demonstrated in whispering gallery microresonators,
such as silica microsphere \cite{Tomes09} and disk \cite{Li13},
crystalline cylinders \cite{Grudinin}. Around the equator of the
microresonators, optical and acoustic waves circulating along the
surface, form ultrahigh quality factor whispering gallery modes (WGMs).
Benefited from the high quality (Q) factor and small mode volume,
the SBS is greatly enhanced when pump, Stocks/anti-Stocks and acoustic
modes are triply on-resonance. This opens new opportunities for coherent
light-acoustic interactions in integrated chips. In last few years,
low threshold Brillouin lasers \cite{Grudinin}, Brillouin optomechanics
\cite{Bahl11,Fluid} and Brillouin cooling \cite{cooling,Tomes11}
have been reported in such triple-resonance WGMs.

\begin{figure*}[tbph]
\centerline{\includegraphics[width=16cm]{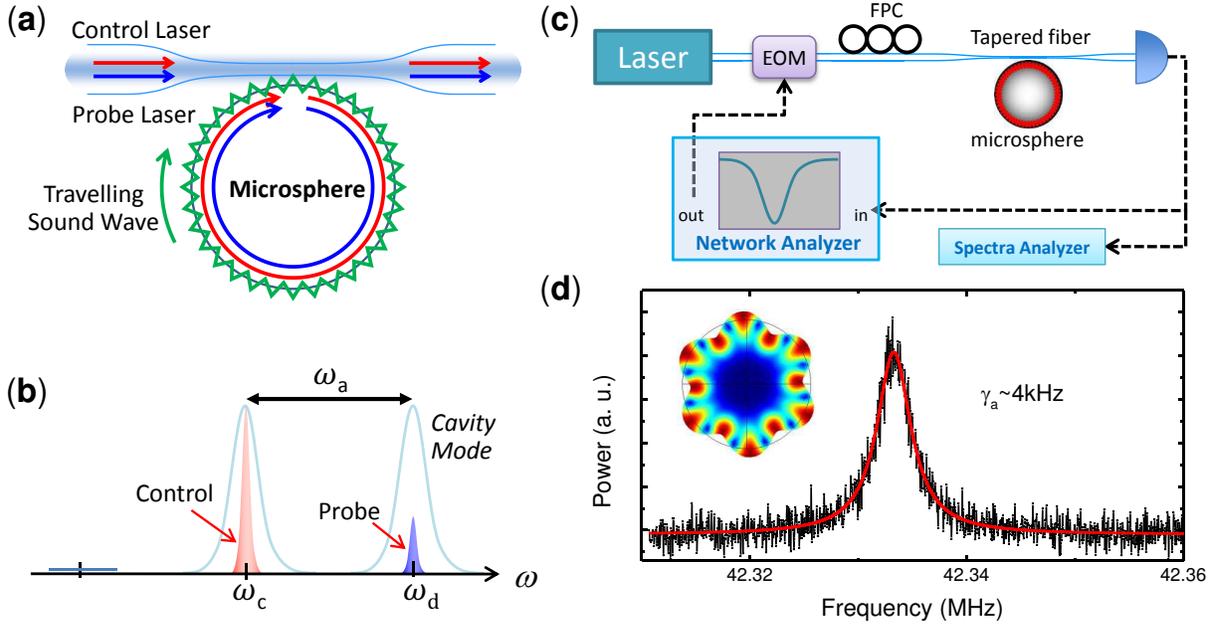}}

\protect\caption{\textbf{Experimental setup of Brillouin scattering in silica microsphere.}
(a) Schematic illustration of the light-acoustic interaction in microsphere.
The optical modes of the microsphere are excited by the pump and probe
lasers through the tapered fiber, and interacting with traveling sound
wave through the forward Brillouin scattering. (b) Energy diagram
of the coherent photon-phonon interaction: pump light is near-resonance
with the lower frequency mode, and the anti-Stocks light as probe
light is on-resonance with another cavity mode while Stocks process
is suppressed. (c) The experimental setup. EOM: electro-optical modulator;
FPC: Fiber polarization controller. (d) The experimental data shows
the typical mechanical mode at $42.3$ MHz in the microsphere when
only pump laser is fixed on-resonance with $\omega_{c}$, the oscillation
is due to the beating between pump and scattered anti-Stocks light.
Inset: Simulation result of the Brillouin scattering mode with $m_{a}=6$
and frequency is $42.2$ MHz.}
\end{figure*}

In this study, we demonstrated coherent Brillouin scattering induced
transparency (BSIT) and non-reciprocal light storage in a silica microsphere
resonator. Placing a strong optical control field on the low frequency
optical WGM, coherent interaction between acoustic and the other optical
WGMs induces a transparency window for the probe light, which is on
resonant with another WGM. Different from the optomechanically induced
transparency (OMIT) that have been observed in a variety of optomechanic
systems \cite{OMIT1,OMIT2}, two optical modes are on resonance with
forward Brillouin acoustic mode in BSIT. Based on the BSIT, a number
of remarkable coherent or quantum optical phenomena are possible,
such as light storage, dark modes and frequency conversion \cite{Dong12,Hill12,Fiore11}.
Especially, we have demonstrated the non-reciprocal light storage
by the SBS, which is potential for quantum memory. These results make
the SBS be great candidate for classical and quantum information processing
in photonic integrated circuits.

In a silica microsphere resonator, there are optical and acoustic
whispering gallery modes (WGMs), which propagate along the surface
{[}Fig. 1(a){]}. Both optical and acoustic WGMs are quantized by the
orbit angular momentum $m$. When the acoustic WGM ($a$) and two
optical WGMs ($c$ and $d$) satisfy the energy and momentum conservations
that $\omega_{a}=\omega_{d}-\omega_{c}$ and $m_{a}=m_{d}-m_{c}$,
photons can be scattered between the optical resonances through Brillouin
scattering \cite{Bahl11}. In this work, we focus on the forward SBS
that $m_{c}$ and $m_{d}$ are with the same sign, and both optical
modes are coupling to tapered fiber, as depicted in Figs. 1(a) and
1(c). As schematically shown in Fig. 1(b), SBS of the pumping on the
optical mode with lower frequency ($\omega_{c}$) leads to phonon
absorption and anti-Stocks photon generation ($\omega_{c}+\omega_{a}$),
while the Stocks process is inhibited. In reverse, the probe light
around $\omega_{d}$ generates phonons and Stocks photons.

We firstly study the SBS by only pump laser around $\omega_{c}$.
We choose a silica microsphere with radius of $98$$\mu$m, and find
a triple-resonance around wavelength $1562$ nm. The scattered anti-Stocks
light is detected by measuring the beating signal between the pump
and scattered light, and the corresponding spectral line is monitored
by an electrical spectrum analyzer, as shown in Fig. 1(d). The Lorentz-shaped
peak indicates an acoustic WGM with frequency $\omega_{a}/2\pi$=
$42.3$ MHz and linewidth $\gamma_{a}/2\pi=4$ kHz ($Q\approx10600$).
The acoustic WGM is also verified by measuring the Stocks scattering
{[}see Fig. S1{]}. From the numerical simulation by finite element
method, the orbit angular momentum of acoustic WGM is identified by
$m_{a}=6$ {[}Inset of Fig.1(d){]}.

\begin{figure*}[tbph]
\centerline{\includegraphics[width=16cm]{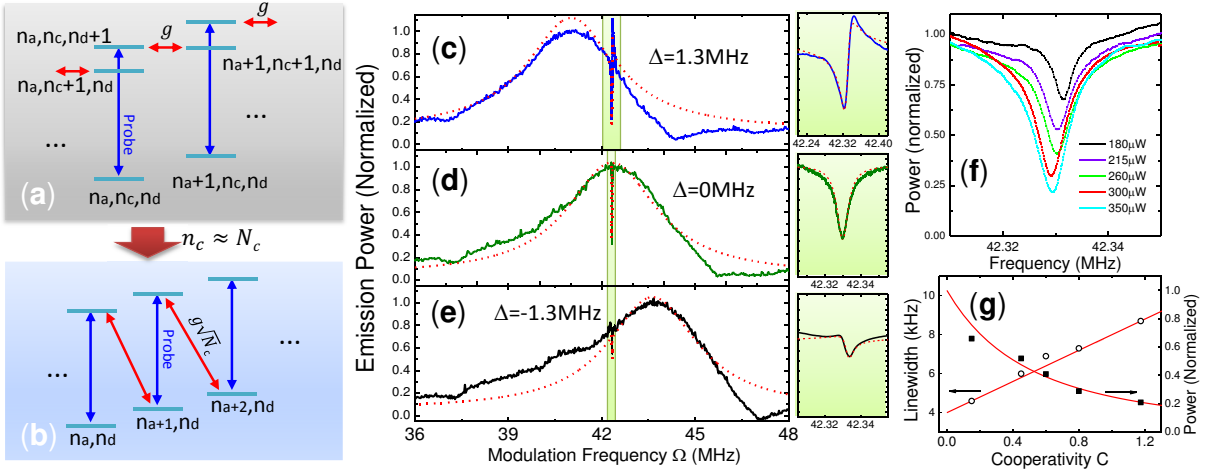}}\protect\caption{\textbf{Mechanism and observation of BSIT.} (a-b) The energy diagram\textbf{\emph{
}}of the triple-resonance in the whispering-gallery cavity.\emph{
}(c-e) Experimentally observed normalized heterodyne traces when the
probe frequency is scanned by sweeping the phase modulator frequency
$\Omega$ for different values of control beam detuning $\Delta=-1.3,\ 0,\ 1.3$MHz.
Whereas the center of the response of the bare optical cavity shifts
correspondingly, the sharp dip characteristic of BSIT occurs always
for $\delta\omega=0$. The input power of the control beam launched
to the cavity is $0.3\ \mathrm{\mu W}$ during the measurements.\ The
insets show the spectral response in the relevant anti-stokes resonance,
with an expanded frequency scale. The short dash lines are the calculation
results with the parameters $k_{d}/2\pi,\ \gamma_{a}/2\pi,\ G/2\pi=3.5,\ 0.004,\ (0.14,\ 0.05,\ 0.03)$
MHz. (f) The emission power from the WGM near the anti-stokes resonance
for four different powers in the control beam from $0.18$ mW up to
$0.35$ mW. The lines are the calculated results. (g) The spectral
linewidth of the BSIT dip and the emission power from the WGM as a
function of the mechanical cooperativity, derived from (f). The solid
lines in (g) show the theoretically expected values for the linewidth
and emission power.}
\end{figure*}

Considering the interaction of forward SBS in the triple-resonance
system, the Hamiltonian can be written as
\begin{align}
H= & \omega_{a}a^{\dagger}a+\omega_{c}c^{\dagger}c+\omega_{d}d^{\dagger}d+g(a^{\dagger}c^{\dagger}d+acd^{\dagger}),
\end{align}
where $a,\ c,\ d$ are Boson operators of acoustic and optical modes
{[}Fig. 1(b){]}. We should note that the vacuum SBS strength $g$
is nonzero only when the three modes are traveling along the same
direction. The energy diagram of the system is\textbf{ }illustrated
schematically in Fig. 2(a), where energy levels are described by phonon
and photon Fock state $\left|n_{a},n_{c},n_{d}\right\rangle $, where
$n_{a(c,d)}$ is phonon (photon) number. The SBS is a parametric process
which induces the transitions between $\left|n_{a},n_{c},n_{d}+1\right\rangle $
and $\left|n_{a}+1,n_{c}+1,n_{d}\right\rangle $, leads to Boson annihilation
and creation of all three modes. The pump and probe lights induce
the transitions that change the photon number of $c$ or $d$ mode,
described by the Hamiltonian $H_{p}=i\sqrt{\kappa_{c,1}}\epsilon_{l}(c^{\dagger}e^{-i\omega_{l}t}-ce^{i\omega_{l}t})+i\sqrt{\kappa_{d,1}}\epsilon_{p}(d^{\dagger}e^{-i\omega_{p}t}-de^{i\omega_{p}t}).$
Here, $\epsilon_{l}$ is strong control laser with frequency $\omega_{l}$
driving on optical mode $c$, and $\epsilon_{p}$ is weak probe light
with frequency $\omega_{p}$ coupling to mode $d$ {[}Fig. 1(b){]}.
$\kappa_{c,1}$ and $\kappa_{d,1}$ are the coupling strength of modes
$c$ and $d$ to the waveguide, respectively.

The vacuum SBS strength $g$ is very weak compared to photon dissipation
rate $\kappa_{c,d}$, thus the $c$ mode is pumped to enhance the
interaction between phonon $a$ and photon $d$, just as people usually
do in the difference frequency generation or sum frequency processes
in nonlinear optical $\chi^{(2)}$processes. For the mean field $\left\langle c\right\rangle =\sqrt{N_{c}}=\left|\frac{\sqrt{\kappa_{c,1}}\epsilon_{l}}{-i\Delta_{c}-\kappa_{c}}\right|$
with $\Delta_{c}=\omega_{c}-\omega_{l}$ , the total Hamiltonian is
simplified in rotating frame as
\begin{align}
H= & -\delta a^{\dagger}a-(\delta+\Delta)d^{\dagger}d+g\sqrt{N_{c}}(a^{\dagger}d+ad^{\dagger})\nonumber \\
 & +i\sqrt{\kappa_{d,1}}\epsilon_{p}(d^{\dagger}-d),
\end{align}
where detuning $\delta=\omega_{p}-\omega_{l}-\omega_{a}$ and $\Delta=\omega_{a}+\omega_{l}-\omega_{d}$.
For a probe laser is near-resonance with $\left|n_{a},n_{d}\right\rangle $
and $\left|n_{a},n_{d}+1\right\rangle $, the model assembles the
well-known electromagnetic induced transparency in $\Lambda$-type
atom. The similar steady state intracavity power spectrum is solved
as
\begin{align}
I_{d}(\delta)\propto\left|\frac{\kappa_{d,1}}{i(\delta+\Delta)-\kappa_{d}/2+\frac{g^{2}N_{c}}{i\delta-\gamma_{a}/2}}\right|^{2}.\label{eq:transmission}
\end{align}
The intracavity power of optical mode $d$ is modified by the coherent
photon-phonon interaction, giving rise to changes of transmission
when $\frac{g^{2}N_{c}}{i\delta-\gamma_{a}/2}$ is comparable to $\kappa_{d}$.
It's convenient to introduce the cooperativity factor $C=\frac{4g^{2}N}{\gamma_{a}\kappa_{d}}$
to evaluate the coherent interaction strength, and $C\gg1$ is preferred
for high fidelity information processing.

To verify the BSIT in our system, we probe the cavity transmission
spectrum in the presence of a control beam. The probe light is generated
from the modulated control laser by EOM, which is modulated at frequency
$\Omega$. The measured total heterodyne signal with the modulation
frequency $\Omega$ using a network analyzer allows extracting intracavity
power, which is directly related to the probe transmission {[}Supplementary
Information{]}. In Figs. 2(c)-(e), we investigate the dependence
of the transparency window on the detuning $\Delta$ by adjusting
$\omega_{l}$ with fixed pumping power $P=300\ \mathrm{\mu W}$. The
shift of transparency window follows the triple resonance condition
when $\delta=0$, and the asymmetric Fano-type lineshapes are in good
agreement with theoretical model {[}Eq. (\ref{eq:transmission}){]}.
We estimate the $C=5.9,\ 0.75,\ 0.27$ for $\Delta=1.3,\ 0,\ -1.3$
MHz. The strongest coupling rate isn't at the detuning $\Delta=0$,
because of the triple-resonance condition is not exactly satisfied
as $\omega_{d}-\omega_{c}-\omega_{a}\approx1.3\ \mathrm{MHz}$. Further
studies of BSIT dip at $\Delta=0$ are shown by Fig. 3(f), for pump
power varying from $180\ \mathrm{\mu W}$ to $350\ \mathrm{\mu W}$.
Dips of increasing depth and width are observed, which can be fitted
by a simple Lorentz function. From Eq. \ref{eq:transmission}, the
expected linewidth is linearly increasing with $C$ relationship as
$(1+C)\gamma_{a}$, and the minimum intracavity power of the BSIT
is given by $\frac{1}{(1+C)^{2}}$. However, we find the $C$ is not
proportional to the input power, due to the thermal effect that changed
the triple-resonance condition.

\begin{figure}[ptbh]
\centerline{\includegraphics[width=7cm]{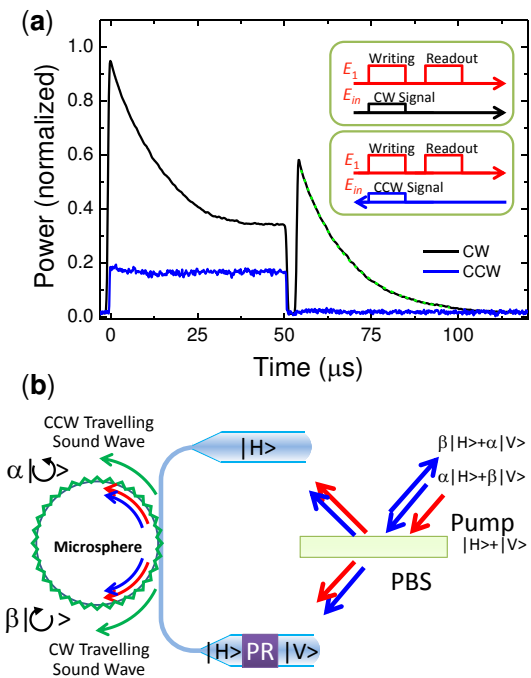}}\protect\caption{\textbf{Non-reciprocal light storage and proposed quantum memory.}
(a) The light storage and retrieval from different input direction.
Inset: the pulse sequence for writing, readout and signal. (b) Schematics
of the single bit quantum memory based on the circulating acoustic
phonons. PBS: polarization beam splitter; PR: polarization rotator.}
\end{figure}

Owing to the coherent SBS interaction, the coherent conversion between
phonon and acoustic phonon could be used for light storage. Especially,
the phase matching conditions of the triple-resonances breaks the
reversal of light propagation. Different from previously studied radiation
and gradient forces driven optomechanics, where mechanical vibrations
could couple to a variety of optical modes, the Brillouin scattering
is only possible for specific mode satisfied energy and momentum conservation
conditions. For example, the input clockwise (CW) pump laser only
permits the interaction between CW acoustic phonon and photon, while
the counter-clockwise (CCW) probe light or acoustic wave is not affected
by SBS. To verify this non-reciprocity, we have studied the light
storage for CW or CCW signal input with fixed CW pump light, as shown
in Fig. 3(a). As evident from the plot, the CW propagating signal
(Black line) is stored and retrieved after $3\ \mathrm{\mu s}$. During
the writing pulse, the measured signal is decreased by the time beacuse
of the dynamical beahvior of BSIT, and similar phenomenon has been
reported of breathing mode in Ref.\cite{Dong_OMIT}. After the writing
pulse, the decay time of the emission power is calculated with $14\ \mathrm{\mu s}$
(Green dashed line in Fig. 3(a)), which is corresponding to the mechanical
linewidth of $15$ kHz. For the CCW launched signal (Red line), there
is no light retrieval during the reading pulse. It's worth noting
that the flat signal in writing pulse is derived from the reflected
signal at the end face of the circulator.

In the previous sections, we have provided experimental evidence of
coherent and non-reciprocal photon-phonon conversion for the SBS in
silica microsphere. Therefore, the degenerated clockwise and counter-clockwise
acoustic modes can be written and read out coherently and independently.
Benefit from these, we proposed a quantum memory of polarization encoded
photon state. As depicted in Fig. 3(b), the horizontal (vertical)
polarized input photon state can couple to CW and CCW optical modes,
store as the CW and CCW acoustic waves and read out, separately. Ideally,
the input state as a superposition of polarization state $\alpha\left|H\right\rangle +\beta\left|V\right\rangle $
can be stored as superposition of circulating acoustic state $\alpha\left|\leftturn\right\rangle +\beta\left|\rightturn\right\rangle $.
As a reversal of the storage, the states are read out by converting
to the photon state $\alpha\left|V\right\rangle +\beta\left|H\right\rangle $
after tens of microseconds.

The Brillouin scattering in the whispering gallery microresonators
is advanced in several aspects. (a) The Brillouin scattering enables
the optical coupling to acoustic wave with frequency range from few
MHz to $11$ GHz, providing a diversified platform for coherent light-matter
interaction. (b) The triple-resonance configure can enhance the SBS,
thus reduce the power consumptions. (c) The traveling wave properties
bring the light non-reciprocity, thus potential for all-optical integrated
non-reciprocal devices, such as isolator and circulator. Our studies
pave the way towards the strong coupling between photon and acoustic
phonon, and encourage further investigations of the non-reciprocity
and memory at quantum level.

\emph{Note:} During the preparation of this manuscript, a similar
work has reported in the Conference on Lasers \& Electro-optics \cite{Bahlcleo}.

\subsection*{Method}

Silica microspheres are fabricated by melting the tapered fiber with
a CO$_{2}$ laser. Optical WGMs in the microsphere are excited through
evanescent field of a tapered optical fiber with a tunable narrow
linewidth ($<300$ kHz) external-cavity laser at the $1550$ nm band.
The experiment setup is schematically illustrated in Fig. 1(c). All
experiments are performed at room temperature and atmospheric pressure.

The coupling strength between the tapered fiber and optical WGMs could
be adjusted by changing the air gap between them, which was controlled
by a high resolution translation stage. The output light is detected
by a low-noise photo-receiver, which was connected to a digital oscilloscope
to measure the transmission spectra or a spectra analyzer to find
the Brillouin scattering modes. For the BSIT experiment, the laser
beam is modulated by the EOM to generate sideband as the probe light.
A network analyzer is used to generate the modulation signal of EOM,
and also measure the spectrum density of beating signal, which corresponding
to the emission power from WGMs.

The measurement of non-reciprocity is carried out by two separated
AOMs, as shown in Fig. S2. A Brillouin mode with $\omega_{a}=152.7$
MHz is chosen to match the working frequency of the acousto-optic
modulators (AOM 1 and 2). The writing and reading pulse array is generated
by AOM 1, where a laser beam is frequency shifted by $-80$ MHz. The
signal pulse is obtained by AOM 2, which is synchronization with the
writing pulse but the frequency is shifted by $+72.7$ MHz. The power
of writing and reading pulses are $1$ mW and the power of signal
is $20$ nW. The duration of writing and reading pulse are $50\ \mathrm{\mu s}$
and $80\ \mathrm{\mu s}$, respectively. The two pulses are separated
by $3\ \mathrm{\mu s}$ as storage time. The forward and backward
signal pulse are coupled into the fiber separately. The power of emitted
light is measured from the spectra analyzer with gate detection mode,
and the resolution bandwidth is $10$ MHz.

\section*{Acknowledgment}

C.H.D. and Z.S. contribute equally to this work. The work was supported
by the National Basic Re- search Program of China (Grant No.2011CB921200),
the Knowledge Innovation Project of Chinese Academy of Sciences (Grant
No.60921091), the National Natural Science Foundation of China (Grant
No.61308079), the Fundamental Research Funds for the Central Universities.

\section*{Author Contribution}

All authors contributed extensively to the work presented in this
paper. C.H.D. and Z.S. prepared microsphere and carried out experiment
measurements. C.L.Z., Y.L.Z. and W.F. provided theoretical support
and analysis. C.L.Z. and C.H.D. wrote the manuscript. C.H.D, C.L.Z.
and G.C.G. supervised the project.

\section*{Competing Interests}

The authors declare that they have no competing financial interests.

\section*{Correspondence}

Correspondence and requests for materials should be addressed to Chunhua
Dong (chunhua@ustc.edu.cn) and Chang-Ling Zou (clzou321@ustc.edu.cn).

\end{document}